\def\ts     {\thinspace}
\def\kms    {\ifmmode{{\rm \ts km\ts s}^{-1}}\else{\ts km\ts s$^{-1}$}\fi}
\def\mo     {M$_{\odot}$}

\documentclass{aa}
\usepackage{graphics} 

\begin{document}
\title{An unbiased deep search for small-area molecular structures}
\author{Andreas Heithausen}
\institute{Institut f\"ur Physik und ihre Didaktik, Universit\"at zu K\"oln, 
Gronewaldstra{\ss}e 2, 50931 K\"oln, Germany}
\offprints {A. Heithausen\\
E-mail: aheithau@uni-koeln.de }
\date {Received 4 January 2006, Accepted 26 January 2006}
\authorrunning{Heithausen }
\titlerunning{Small-area molecular structures}

\abstract
{Small-area molecular structures (SAMS) resembling those
clumpuscules, proposed by Pfenniger and Combes (1994) as candidate for
baryonic dark matter, have recently been detected (Heithausen 2002, 2004) in an
area where the shielding is too low for them to survive for a long time.}
{To study the frequency of occurence of such structures 
I present the results of an unbiased deep search for molecular
clumpuscules. } 
{The area surrounding these structures has been surveyed using the FCRAO
14m telescope in the CO $(1\to0)$ transition. The field covered is 20$'$ by
20$'$. The resulting rms of the data is only 0.04\,K in a 0.127\,\kms\ wide
channel. Also, high-angular resolution observations of the $^{13}$CO and
C$^{18}$O $(J=1\to0)$ transitions were obtained with the IRAM Plateau-de-Bure
Interferometer.}
{3 new SAMS have been detected.  The structures
have very low intensities which makes it impossible to detect them in large
scale CO surveys conducted to map the molecular gas of the Milky way.  They
move with a similar radial velocity as the surrounding HI gas.  
The clouds follow the
same size-linewidth relation as found for giant molecular clouds or Galactic
cirrus clouds. The observations clearly show that most of the large
linewidths
observed at low angular resolution is caused by a large velocity difference
among the clumps seen at highest angular resolution.

The non-detection of the
structures in the high-angular
resolution observations of the $^{13}$CO and C$^{18}$O $(J=1\to0)$ transitions
shows that the $^{12}$CO $(J=1\to0)$ transition must have a low optical depth.
At an adopted distance of 100\,pc the structures have masses of only Jupiter
mass or below. }
{My observations show that SAMS might be an abundant phenomenon in the
  interstellar medium however not recognized as such due to their small size.
If they are made of ordinary interstellar matter with solar metallicity
they likely contribute only little to the total interstellar mass.}
  \keywords{Dark matter - Interstellar medium (ISM): abundances - ISM: clouds 
- ISM: molecules}         
\maketitle

\section{Introduction}

It has been proposed that most of the baryonic dark matter in the Galaxy could
be made of small, dense clumps of molecular hydrogen
 (Pfenniger \& Combes \cite{pfenniger:combes1994}; de Paolis et al. 
\cite{depaolis1995}; Gerhard \& Silk \cite{gerhard:silk1996}; Walker \&
Wardle \cite{walker:wardle1998}). The detection of H$_2$ is notoriously 
difficult due to the missing
emission lines of that molecule at temperatures below a few hundred K.
Sensitive absorption line studies towards many lines of sight have shown that
H$_2$ is ubiquitous in our Galaxy (e.g. Shull et al.  \cite{shull:etal2000};
Richter et al. \cite{richter:etal2003a}; Jenkins et al. \cite{jenkins:etal2003})
even in the diffuse medium.

While absorption line measurements indicate that there is significant
small-scale structure even in the molecular gas, they do not provide much
information on the structure or the form of the absorbing clouds. The recent
detection of small-area molecular structures (SAMS) in emission of the CO
$(J=1\to0)$ transition (Heithausen \cite{heithausen2002}, hereafter Paper 1)
thererfore provided the first chance to study their small scale structure in
detail. The detection was somewhat puzzling because the clouds were observed
in an area where the shielding is too low for them to survive for a longer
time.

Follow-up high-angular resolution observations (Heithausen
\cite{heithausen2004}, hereafter Paper 2) showed that the structures
resembled those clumpuscules predicted by Pfenniger \& Combes
(\cite{pfenniger:combes1994}) as important candidates for the baryonic dark
matter, albeit warmer and probably less massive. These observations also
revealed that all the mass seen on larger scale is contained in smaller
structures, a few hundred AU large, with no need for a diffuse molecular
component.

Here I present the results of a deep search for further small-area molecular
structures in the area where the first ones have been detected. I will show
that these structures are not isolated objects but group in small clusters.
Furthermore I will discuss high-angular resolution observations of SAMS2 in
the $^{13}$CO and C$^{18}$O $(J=1\to0)$ transition which provide information
on the optical depth in the same transition of the main isotopomer $^{12}$CO.
These information will allow a better estimate of the mass of the objects and
thus of their relative importance. 

\section{Observations}

\begin{figure}
\rotatebox{-90}{\resizebox{6.5cm}{!}{\includegraphics{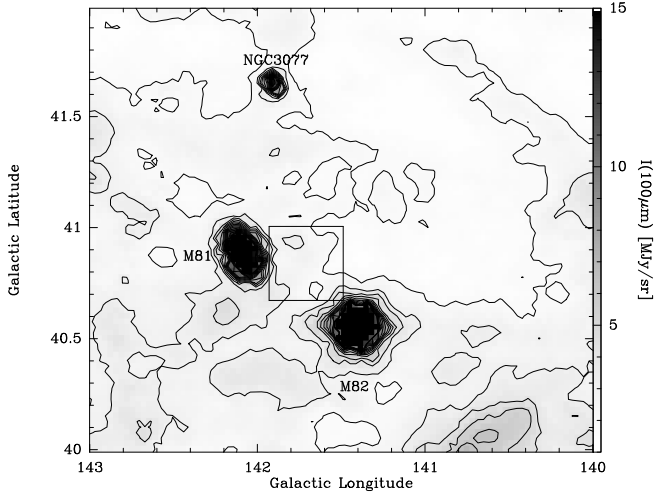}}}
\caption{IRAS 100 $\mu$m map of the area surrounding the field observed
 in the CO $(1\to0)$ line with the FCRAO 14m telescope (marked by the
square). The intense infrared point sources M81, M82, and NGC3077 are labelled.
Contours are in steps of 1\,MJy sr$^{-1}$ starting at 1\,MJy sr$^{-1}$.}
\label{overview}
\end{figure}

\subsection{FCRAO observations}

To study whether SAMS2 is an isolated object, a large-scale map
(see Fig. \ref{overview}) in the CO $J=1\to0$ transition was obtained with the
14m telescope of the Five College Radio Astronomical Observatory (FCRAO)
near Amherst in Massachusetts, U.S.A., in April 2003 and
January and February 2004.  Using the Sequoia 32 pixel receiver, 
data were taken in the on-the-fly mode with an
off-position adopted to be free of emission about 1 degree away from the
source position. The area was sampled with a spacing of $20''$ between
individual positions. The angular resolution of the 14m telescope at 115\,GHz
was $44''$, the main beam efficiency was 0.45.  The resulting map has a size
of $20'$ by $20'$. The spectra were sampled with a velocity resolution of
0.127\,\kms; the resulting rms in the spectra was 0.04\,K.

\subsection{Plateau-de-Bure observation}

High angular resolution $^{13}$CO and C$^{18}$O data in the $J=1\to0$
transition were obtained with the Plateau-de-Bure interferometer near Grenoble
in the French Alps between May and December 2004.  The observations were
conducted in the C and D configuration which gives an angular resolution of
$3''$ by $3''$.  The map consists of 6 individual pointings covering an area a
bit larger than that observed in the $^{12}$CO $(1\to0)$ transition
(Paper 2). The final spectral resolution was 0.1 \kms\
at 110GHz.

\section{Results \label{results}}

\subsection{CO on larger scale}

\begin{figure*}
\rotatebox{-90}{\resizebox{16cm}{!}{\includegraphics{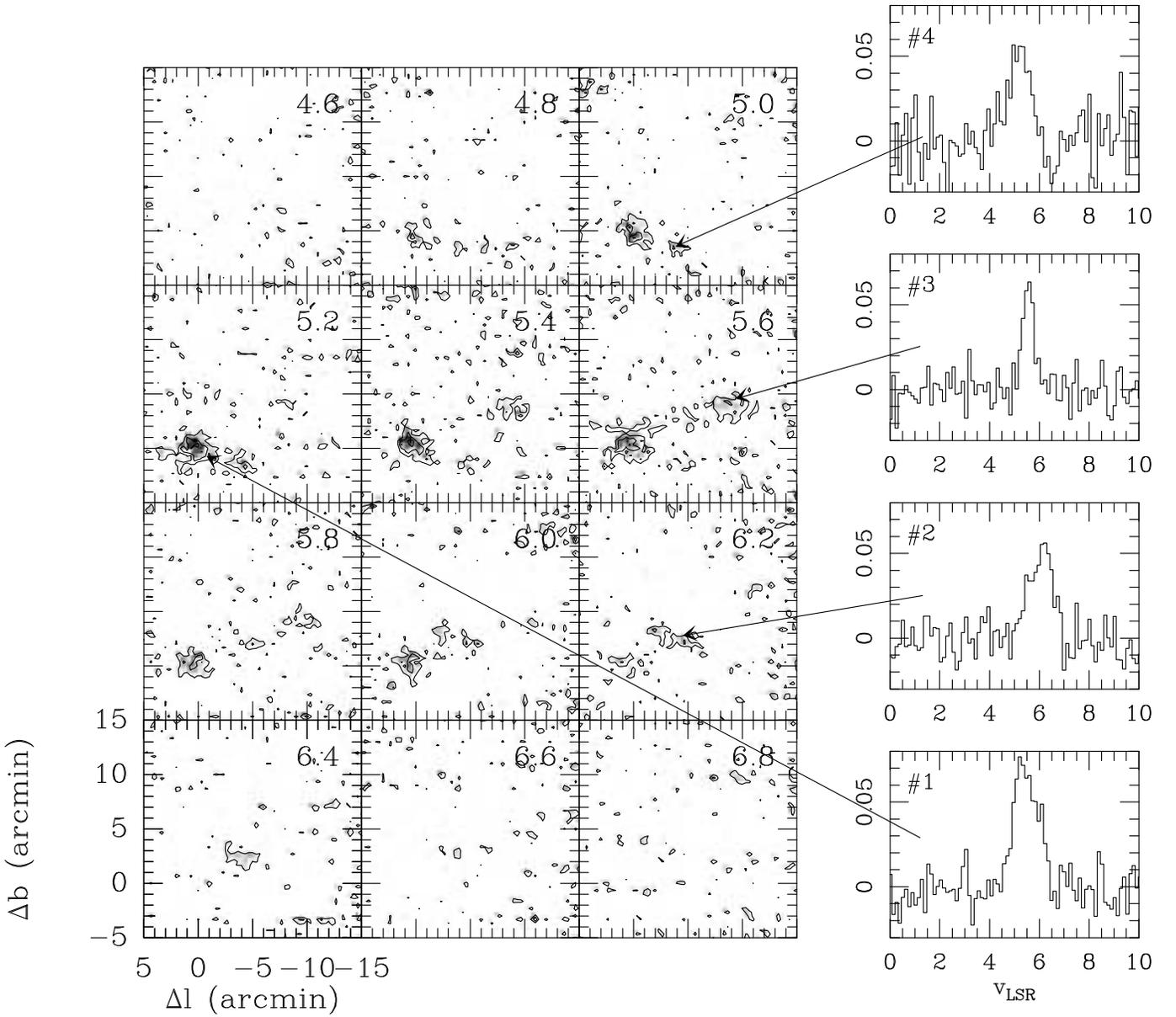}}}
\caption{Channelsmaps of the large-scale CO observations obtained with the FCRAO
14m telescope. The velocity with respect to the local standard of rest of each
channel is indicated in the upper right corner of each map. Offset in Galactic
coordinates are relative to $l_0=141^\circ\!.71, b_0=40^\circ\!.85$. Contours are
every 0.06\,K\ starting at 0.06\,K\ ($2\sigma$). To the right spectra averaged
over the individual substructures are displayed. Temperature scale is
$T_{\rm A}^*$\ and the velocity $v
_{\rm LSR}$\ in \kms.}
\label{fcraochan}
\end{figure*}

Fig. \ref{fcraochan} shows the results of the unbiased observations of a
large area around SAMS2. Presented are maps integrated over 0.2\,\kms\ wide
velocity channels. In these maps four individual clouds can be found: \#1 
corresponds to SAMS2, the other three have previously not been identified. 
There are even more candidates for molecular clouds in the field which are
however close to the $3\sigma$ detection limit and thus have to be
confirmed. Most of those are found in the velocity range between 5 and 7 \kms\
where the other clouds are detected. This velocity range is identical to that
of the HI gas in the same direction.
 
Note that with the on-off observing method chosen for the observations, signals
which change not much between the on and off position 
in both velocity and intensity are filtered out. Due to
the location of the off-position at $(-60',0')$ from the $(0,0)$ position of the
map $(l_0=141^\circ\!.71, b_0=40^\circ\!.85)$ diffuse 
clouds larger than 1 degree cannot
be detected. Clouds appearing in on- and off-position with velocities differing
by more than the velocity resolution (0.127\kms) would show up as P-Cygni type
line profiles with clouds in the off-position as apparent absorption dip. Such
profiles were however not detected in the data.

Values from a gaussian fit to the cloud averaged spectra are listed in
Tab. \ref{fcraotable}. Line parameters are derived from spectra averaged over
the area with significant emission. Radii $r$ given in that table were
determined from that area $A$ via $r=\sqrt{A/\pi}$.  The center positions were
determined from a two dimensional gaussian fit to the clouds.  

\begin{table*}
\caption{Parameters for the molecular structures obtained from the FCRAO data}
\begin{tabular}{l l l l l l l l l l}
\noalign{\hrule}
\noalign{\medskip}
 Number & $l$ & $b$ & 
$T_{\rm A}^*$ & $rms$ & $v_{\rm lsr}$ & $\Delta v$ & Radius$^{(1)}$ 
& $M_{\rm H_2} ^{(2)}$ & Identification\\
        &  (deg) & (deg)    & 
(K)     &  (K)  & (\kms)    & (\kms)     & ($'$) & 
$10^{-3}\,({d\over100{\rm pc}})^2$\mo\\
\noalign{\medskip}
\noalign{\hrule}
\noalign{\medskip}
\#1 & 141.723 & 41.860 & 0.072 & 0.010 & $5.45\pm0.04$ & $1.13\pm0.08$ & 1.9
& 1.4 & SAMS2\\
\#2 & 141.666 & 40.895 & 0.053 & 0.011 & $6.07\pm0.05$ & $1.07\pm0.10$ & 1.6
& 0.7 & $-$ \\
\#3 & 141.518 & 40.922 & 0.060 & 0.012 & $5.57\pm0.03$ & $0.52\pm0.07$ & 1.5
& 0.3 & $-$ \\ 
\#4 & 141.615 & 40.822 & 0.055 & 0.015 & $5.19\pm0.07$ & $1.00\pm0.16$ & 0.9
& 0.2 & $-$ \\
\noalign{\medskip}
\noalign{\hrule}
\noalign{\medskip}
\noalign{Remarks: 1) the radius $r$\ is determined from the area $A$\ 
covered by spectra with emission via $r=\sqrt{A/\pi}$. 2) Masses are pure
H$_2$ masses not corrected for the contribution of Helium; they are calculated
using the X-factor. Masses determined assuming a low optical depth are a
factor 13.6 lower (see text).}
\end{tabular}
\label{fcraotable}
\end{table*}

\subsection{CO excitation and cloud masses\label{excitation}}

In the high-angular resolution $^{13}$CO and C$^{18}$O $(J=1\to0)$
observations no signal of more than 0.3 K ($3\sigma$) was detected compared to
lines up to 6 K in the same velocity range of the $^{12}$CO line (Paper 2). 
In typical Galactic molecular clouds both
the $^{12}$CO and the $^{13}$CO $(J=1\to0)$ lines are readily detected with an
line ratio of ${T({\rm ^{12}CO})}/{T({\rm ^{13}CO})}\approx 3-5$
(e.g. Falgarone et al. \cite{falgarone:etal1998}), much higher than one would
expect from the isotopic ratio $[^{12}{\rm C}]/[^{13}{\rm C}]=66$ (e.g. Bensch
et al. \cite{bensch:etal2001}) under the assumption that the lines are
optically thin. It is therefore normally concluded that (at least) the
$^{12}$CO line is optically thick. For SAMS2 the intensity ratio is much
higher ${T({\rm ^{12}CO})}/{T({\rm ^{13}CO})}\ge 20$, which means that the
$^{12}$CO line has a significant lower optical depth than the same line in
other Galactic molecular clouds.

The H$_2$ masses of the structures were therefore determined in two ways:
\begin{itemize}
\item adopting a CO to H$_2$ conversion
factor of $X_{\rm CO}=1.5\times 10^{20}$ cm$^{-2}$ (K \kms)$^{-1}$, i.e. the
Galactic value (Hunter et al. \cite{hunter1997}),
\item adopting optically thin CO emission and a CO abundance of $N_{\rm
CO}/N_{\rm H_2}=1.0\times 10^{-4}$ (Sutton et al. \cite{sutton:etal1995}). 
\end{itemize}

Using the X-factor and adopting a distance of 100\,pc to the clouds 
I derive masses in the range of Jupiter masses $(M_{\rm
J}=0.001M_\odot)$.  The mass for SAMS2 determined this way is in agreement
with that determined from the IRAM 30m data (Paper 1) given the uncertainties
in the determination of the radii and the respective main beam efficiencies of
the telescopes used.

Assuming optically thin emission and adopting a excitation temperature $T_{\rm
ex}=20$\,K (see Paper 2) results in H$_2$ masses which are a factor of 13.6
lower than those determined with the X-factor.  Note that the CO abundance in
regions with low shielding is expected to be lower (e.g. van Dishoeck \& Black
\cite{vandishoeck:black1988}) than the value for large molecular clouds given 
by Sutton et al.; the masses
determined that way are therefore lower limits to the real value (provided the
distance is correct).

\section{Discussion}

\begin{figure}
\rotatebox{-90}{\resizebox{7.5cm}{!}{\includegraphics{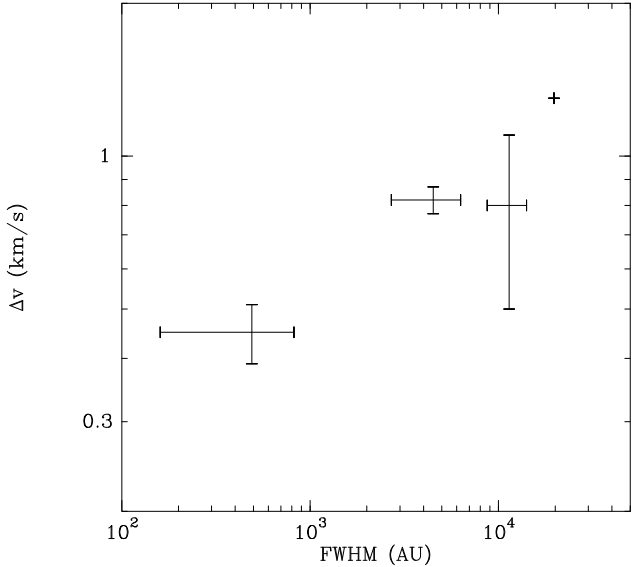}}}
\caption{Size-linewidth relation for SAMS. The correlation includes data also
  from Paper 1 and 2. Errorbars represent the range of values from the various
  data sets rather than formal errors. }
\label{sizelinewidth}
\end{figure}

\subsection{Velocity}

The observations presented in this paper show clearly that SAMS break up in
more and more clumps when going from low to high angular
resolution. E.g. SAMS2 seen with the FCRAO telescope shows up as a single
structure with indication for further substructure. Observed with the IRAM 30m
telescope at least two substructures are visible (Paper 1). At high angular
resolution observed with the PdB interferometer at least 6 smaller stuctures
are observed, some of which are even unresolved (Paper 2).  While the angular
resolution increases the line of sight line width decreases from about
1\,\kms\ at a resolution of 44$''$ to about 0.4\,\kms\ at 3$''$.  It is
obvious that most of the large line width at low angular resolution is caused
by a large velocity gradient among the smallest visible structures.

I therefore first analyse the size-linewidth relation which provides a good
estimate for the amount of turbulence.  Before however we use the data from
Paper 1 and 2 to compare them with the data presented in this paper we have to
correct for the different definition of the sizes. In Paper 1 and 2 I defined
the sizes as full width at half maximum $FWHM$ of the structures derived from
a gaussian fit to the data. In this paper the radius $r$ was derived from the
total area covered by significant emission. The latter definition depends
therefore on the signal-to-noise ratio $S$ of the data, i.e. a higher $S$
results in a larger size. The correction factor can easily be calculated from
a comparison of their different definitions: $ {\sqrt{-2{\rm ln}(S)}\over
2.355}FWHM=r$. For the actual $S=1/5$ it follows that $r=0.76 FWHM$.

The  data corrected that way
 are plotted in Fig. \ref{sizelinewidth}. This figure also
includes a single value for the whole ensemble of clouds in the area; for that
the radius has been calculated from the total area covered by significant
emission, and the linewidth from the sum of the spectra shown in
Fig. \ref{fcraochan}.  Rather than plotting formal error bars the figure shows
the range of values for the different scales observed. The data nicely follow
a relation of the form
$$\Delta v \propto FWHM^\alpha$$ 
with $\alpha=0.3\pm0.1$ which is within the range of values found for giant
molecular clouds (Larson \cite{larson1981}) and Galactic cirrus clouds
(Heithausen \cite{heithausen1996}).

\subsection{Abundance \label{abundance}}

Even with the deep observations presented in this paper it is hard to
determine the relative importance of the small molecular structures.
Due to their very low sizes it is clear that we only see those which are
very close to us and bright. At a distance three times larger than their
actual value only the structure \#1 (SAMS2) would have been detectable with the
FCRAO telescope, however just in one single spectrum with very low intensity.
So if the distance is indeed 100pc (see Paper 1), we can
only see SAMS in a volume of 300pc radius with the currently available
radiotelescopes. If SAMS are however a widespread phenomenon they should be
easily seen even at larger distances once ALMA comes into operation.

The analysis of the relative importance of SAMS relies furthermore on the
assumption that important physical or chemical 
parameters, as e.g. the CO abundance or the kinetic
temperature, are similar to that found for larger molecular clouds; an
assumption which not necessarily needs to be true. As discussed in
Sec. \ref{excitation} we possibly overestimate the CO abundance and thus
underestimate the masses and column densities of SAMS. 
Dirsch et al. (\cite{dirsch:etal2003}, \cite{dirsch:etal2005}) recently
reported the detection of a tiny dust cloud of 4$''$ diameter seen in
projection towards the spiral galaxy NGC\,3269. They derive a maximum
absorption through the cloud of $A_{\rm B}\approx 1$\,mag. If this dust cloud
is also associated with molecular gas it could turn out to be the first
optically detected SAMS. In that case a more reliable value of the column
density and mass could be derived which is significantly higher than that
estimated for the SAMS in this paper. 

Averaged over the whole 20$'$ by 20$'$ area the CO line has an integrated
line strength of $W_{\rm CO}= 0.011\pm0.002$\,K\kms. The CO clouds thus 
account for a molecular
hydrogen column density of $N({\rm H_2})=(0.0012-0.017)\times
10^{20}$\,cm$^{-2}$; here the lower value is calculated adopting optically
thin emission and a Galactic CO abundance and the upper value adopting a
Galactic $X_{CO}$ factor (see Sec. \ref{excitation}). The HI column density of the
same area already given in Paper 1 is $N({\rm HI})=2.1\times
10^{20}$\,cm$^{-2}$. That means that only 0.06-0.8 percent of the total gas in
that area are in form of molecular gas if the assumptions made are correct.

\section{Conclusions}

The deep observations presented in this paper have more than doubled the
number of known SAMS. As already discussed in Paper 1 their velocities are all
within the same velocity range as the local HI clouds towards that direction
of the sky. They therefore are probably a natural constituent of the
interstellar medium, however not recognized as such due to their small
angular sizes and low intensities.  While on smallest angular scales the
molecular gas of SAMS dominates the total amount of the interstellar gas in
the corresponding velocity range (as shown in Paper 2) on larger scales they
form possibly only a minor fraction.  This result relies
however on the assumption that SAMS are made of ordinary interstellar
matter. If they are made of pristine material, only slightly enriched with
metalls, SAMS could be more massive, because we overestimate the CO abundance
significantly (see Sect. \ref{abundance}). In this case they could even be
virialized, making them more massive, 
similar to the clumpuscules proposed by Pfenniger
\& Combes (\cite{pfenniger:combes1994}). Consequently, they would also
contribute more mass to the interstellar medium.  Only direct observation of
H$_2$ lines, e.g. in absorption to distant stars or quasars, could help to
solve this issue.

The detection of Galactic H$_2$ in an HI high-velocity cloud let Richter
et al. (\cite{richter:etal1999}) conclude that the cloud may be dissolving.
H$_2$ clouds have now been seen towards many extragalactic lines of sight (see
Richter et al. \cite{richter:etal2003a},
\cite{richter:etal2003b}, Shull et al. \cite{shull:etal2000}).
Some of these clouds have similar physical properties, i.e. high densities and
small sizes, as the CO clouds dicussed here.  The high detection rate suggests
that these are not all in the state of dissolution, but rather that there is a
steady process of formation and destruction of H$_2$ clouds in the
ISM. Because with respect to fractal index and size-linewidth relation SAMS
are similar to larger molecular clouds, the observations presented here and in
the previous papers also suggest that the structural properties of molecular
clouds are imprinted in right in the beginning of their formation process.
High-angular resolution HI data, to compare with the CO data, are necessary to
shed further light on this issue. Such data are currently being reduced and
will be presented in a forth-coming paper.

\begin{acknowledgement}
I thank Arancha Castro-Carrizo (IRAM) for her help during the data reduction
of the PdB data and Philipp Richter for comments on the manuscript.
This paper is based on observations with the Five College Radio Astronomical
Observatory (FCRAO) and the IRAM Plateau-de-Bure interferometer. IRAM is
supported by INSU/CNRS (France), MPG (Germany), and IGN (Spain).
\end{acknowledgement}

{}

\end{document}